\documentstyle[twocolumn,prl,aps,floats]{revtex}

\begin{document}

\input{psfig.sty}

\wideabs{
\title{The Los Alamos Spallation Driven 
Solid Deuterium Ultracold Neutron Source: 
Working Notes of Mar. 30, 1998 with Embellishment}

\author{S.K. Lamoreaux}

\address{University of California,
Los Alamos National Laboratory, Physics Division, Los Alamos, NM 87545}

\date{\today}

\maketitle
\begin{abstract}
The working notes
which led to the physics demonstration ultracold neutron source that
has been operated at Los Alamos are transcribed here.  
In addition to the transcribed
notes, included are a Prologue that describes the path that led
to the basic idea, and an Epilogue that describes some of
the discoveries
in the implementation of the idea.
\end{abstract}} 

\section{Prologue}

The idea that solid deuterium could be a useful source of ultracold
neutron (UCN) dates to the early 1980's
when R. Golub {\it et al.} [1] proposed the ``thin film''
UCN source.  The basic idea is that a material, such as solid
deuterium (SD$_2$) has a finite neutron absorption cross section,
but a high rate of UCN production compared to superfluid helium.
A small amount of SD$_2$  contained in a UCN storage cell and
irradiated by cold neutrons will produce UCN until the rate of
production equals the rate of absorption.  Because both processes
are proportional to the amount of SD$_2$, the steady-state
UCN density is independent of the amount of material. Although the
production rate is higher for SD$_2$ than for superfluid helium, 
the ultimate density of UCN in a superfluid helium filled long 
lifetime storage cell is around 10-100 times greater.

However, there are applications where SD$_2$ will be useful for
a UCN converter.  It has been suggested that the thin film source
would be easier to use near the core of a high-flux reactor, primarily
because less material is required to achieve a particular UCN current.

A specific application of SD$_2$ to UCN production at a spallation
neutron source was brought to our attention in Jan. 1998 by A. Serebrov [2]
at a workshop held near St. Petersburg, Russia.
The basic idea is that the UCN storage bottle need only be connected
to the SD$_2$ during a short time after the spallation pulse; one
can then enjoy a high production rate together with an effectively
increased storage lifetime.  Upon returning to
Los Alamos we (D. Bowman, S. Lamoreaux, C. Morris)
discussed the ``UCN Factory'' described
in [2] and came to the conclusion that it had a number of problems
that include wasting phase space of the cold neutron flux, the necessity
of a UCN window between the SD$_2$ and storage vessel, the use of an unwieldy
amount of SD$_2$ (approaching 1 m$^3$), an overly complicated
shutter mechanism, and it seemed that the 
SD$_2$ thickness was such that UCN might not be able to propagate
out of the material irrespective of the window.  Such a system
might eventually be made to work, but without understanding the
basic physical mechanisms involved, it seemed unwise to invest the large
amount of resources required to build even a scaled version of
the Factory.

In addressing these problems, in particular the cold neutron phase space,
we were challenged as to why they were indeed problems.  In answering
the challenge, the following model source was proposed.  
Upon further reflection,
the geometry is close to ideal for our specific application.

Our model source closely follows a proposal by Pokotilovski [3] who
really came up with the idea of separating the production and
storage regions, with a shutter that opens only around the
time of the spallation pulse.  R. Golub had provided S. Lamoreaux
with a copy of this paper in 1995, but we had no conscious memory
of this paper at the time of the St. Petersburg workshop; the paper
emerged when we were digging through our notes.  Our
model source has only minor embellishments over the original
Pokotilovski proposal which the interested reader can find
by comparing these notes to [3].

\section{Notes of Mar. 20, 1998}

\subsection{Basic scheme of source}

A rough schematic of the physics demonstration UCN source is
shown in Fig. 1.

\begin{figure}
\centerline{\psfig{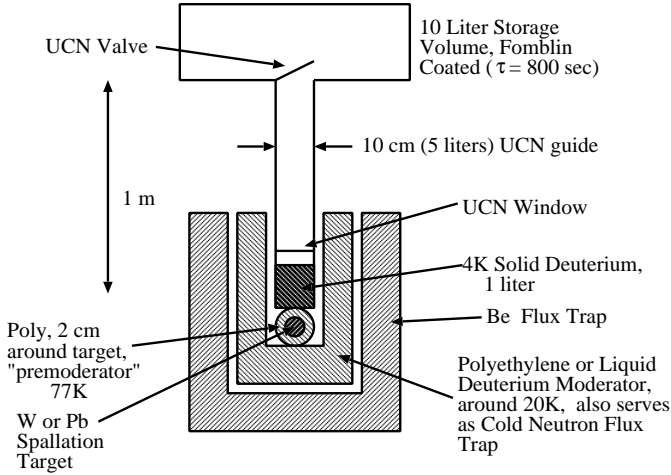}}
\caption{Working model (not to scale) 
of a solid deuterium spallation driven UCN source; the UCN valve
closes after a short spallation pulse,
at the time of maximum UCN density in the upper storage
chamber.  The 1 m vertical rise
compensates the solid deuterium material potential (about 100 neV).}
\end{figure}
\subsection{Cold neutron production:}

15 neutrons/800 MeV proton from spallation, 
\begin{equation}
P_c=15{I\over e}\approx 10^{14}/{\rm sec\ at\ 1}\mu{\rm A}
\end{equation}
($P_c$ is the cold neutron production rate).

If the poly or LD$_2$ moderator is 30 K or lower, the produced neutrons
will moderate to $\sim$30 K (moderation processes turn off for
neutron temperatures below $\sim$25-30 K).

For poly, the albedo is about 90\% for 30 K neutrons.  We can estimate
the mean free path of a cold neutron in the flux trap region as
$\approx$ 40 cm before loss (up the UCN guide) or absorption.  This gives
a cold neutron dwell time in the flux trap region
\begin{equation}
\tau_c={0.4\ {\rm m}\over 700\ {\rm m/s}}=0.66\ {\rm ms}.
\end{equation}
The cold neutron density in the flux trap region is them
\begin{equation}
{P_c\tau_c\over V_c}=\rho_c.
\end{equation}
Take $V_c$ as 1.5 times solid D$_2$ volume $V_D$ 
\begin{equation}
V_c=1.5 \ell.
\end{equation}
So, at 1 $\mu$A,
\begin{equation}
{10^{14}\over {\rm sec}}\cdot 0.66\times 10^{-3}{\rm sec}\times 
{1 \over 1500 {\rm cc}}= 4.4\times 10^{7}/{\rm cc}
\end{equation}
Then, the cold neutron flux is 
\begin{equation}
{1\over 4}\rho_cv_c; \ {\rm with\ v_c=700\ m/s}.
\end{equation}
\begin{equation}
\Phi_0=8\times 10^{11} {\rm n/cm^2 s}
\end{equation}

\subsection{UCN Production rate}

See Fig 3.17 of {\it Ultracold Neutrons} (Golub, Richardson,
Lamoreaux) [4].  For the thin film source, 
$\rho=150\times 10^{-11}\Phi_0/{\rm cm^3}$.

The production rate (of UCN) is $P_u=\rho/\tau$ and $\tau=0.12 {\rm sec}$ at
$T<4$ K.  So, $P_u=1.25\times 10^{-8}\Phi_0$ for SD$_2$ in a 
Be bottle.  We must correct for the potential; $U_{SD_2}=108$ neV,
this is compensated by the vertical rise.  For the thin film
source, $P\propto (U_{Be}^{3/2}-U_{SD_2}^{3/2})$ for our
source, $P' \propto U^{3/2}_{Fomblin}$
\begin{equation}
{P\over P'}={252^{3/2}-108^{3/2}\over 160^{3/2}}=1.42
\end{equation}
so four our system $P_u=9\times 10^{-9}\Phi_0$.

Total number of UCN produced is
\begin{equation}
V_D\cdot P_u=1000\cdot 9\times 10^{-9}\cdot 8\times 10^{11}
=7.2\times 10^6 {\rm UCN/sec}.
\end{equation}

\subsection{UCN lifetime and UCN density}
We have
\begin{equation}
{7.2\times 10^6{\rm UCN/sec}\over 15000 {\rm\ cc\ total}}=480 {\rm UCN\over 
sec\ cm^3}.
\end{equation}
The lifetime is given approximately by the ratio of the total volume
$V_T$ to the solid D$_2$ volume, times the UCN lifetime in solid
D$_2$.
\begin{equation}
{V_T\over V_D}\tau_D={15\over 1}\cdot 0.12 {\rm sec}=1.8\ {\rm sec}.
\end{equation}
This gives 
\begin{equation}
\rho_{ucn}={\rm 480\ UCN\over cm^3 s}\cdot 1.8 {\rm s}=764 {\rm\ UCN/cm^3}.
\end{equation}
\subsection{Equilibration time etc.}

The equilibrium time for the system is 1.8 sec, the shortest lifetime
in the system. Thus, the 1 $\mu$A beam would have to be turned
on for 3.6 sec (two lifetimes) after which the shutter to the
storage volume is closed.

After 3.6 sec of filling, the storage volume will be filled with
$\approx 10^7$ UCN.  If these are released at a constant rate over
400 sec, loss will be minimized, and the UCN current will be
$10^7/400=2.5\times 10^4$/sec.

The 800 MeV beam would have to be applied every 400 sec; this rate
could be increaed for a higher UCN flux.  Also, the 800 MeV beam current
could be increased; it's a question of radiation damage and heat
load.  (Power for 1 $\mu$A, 800 MeV=800 Watts.)

\section{Epilogue}

The prototype source was constructed and tested in Aug. 1998.  As
anticipated, the window created a significant UCN loss; {\it e.g.},
there is an effectively small shunt impedance in parallel with the
source, as shown in Fig. 2. 

\begin{figure}
\centerline{\psfig{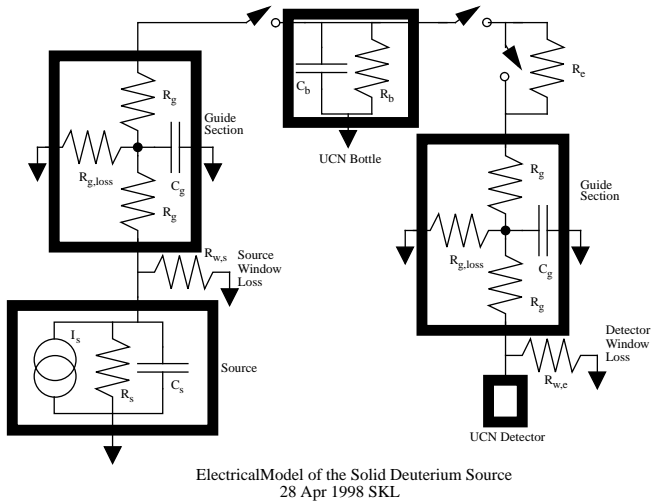}}
\caption{Electrical analog of the solid deuterium source
shown in Fig. 1.  The shunt impedance offered by
the window is very small thereby sinking UCN away from
the guide, as has been the case of
many cryogenic UCN source experiments. This window was
eventually removed.}
\end{figure}

A windowless
system was designed and built; yet, a substantially lower UCN density compared
to these notes was obtained.  This was due to a number of causes.

The state of affairs as of late 1998 are given in [5].

First, the flux trap is rather more complicated that the simple
estimate above would indicate.  The cold flux density extends
further into the moderator and Be reflector, so the mean free
path is modified. Monte Carlo calculations by R. Hill indicate
that the flux trap lifetime is closer to 160 $\mu$ sec, and
the effective volume is larger.  This results in a
factor of 10 reduction in cold neutron flux.

In a test of the production rate of UCN by scattering cold
neutrons in solid deuterium in March 1998, done at the Hahn-Meitner
Institut in Berlin, the rate was within a factor of 3-5 of the
expected rate given in Sec. II.C  above, assuming all
detected neutrons were UCN.  However, the expected temperature
dependence to the apparent production rate was not observed
(possibly due to loss of UCN before they can propagate out
of the solid).  The observed UCN flux and lack of temperature
dependence could be explained by assuming a UCN lifetime of
a few milliseconds in the solid.
The same lack of temperature dependence was initially observed
in the demonstration source; we eventually found that the para
molecular state, which persists from its room temperature concentration
through freezing at 4 K, can upscatter UCN with a lifetime of
about 4 milliseconds [6].  

A para/ortho converter was installed on the demonstration
source and the observed UCN production rate agrees well with
Monte Carlo calculations, and a temperature dependence is observed.

The hydrogen contamination of available deuterium gas (0.1 at\%),
upscattering by phonons at 5 K (lowest practical temperature),
and 2\% para contamination (lowest practical conversion temperature
is 17 K at which temperature there is still a reasonable vapor pressure),
gives a net lifetime of 25 milliseconds.

A final weakness in the working notes is the assumption that one
liter of solid D$_2$ could be used; in fact, the mean free path
of a UCN is about 8 cm due to the incoherent
cross section alone (and is possibly shorter for 
an amorphous material) and limits the volume; roughly 200 cc
is a practical upper limit for the solid volume for the geometry
shown in Fig. 1.

These factors taken together give a factor of 10-100 less UCN
than the simple calculation.  Nonetheless, the concept still has
potential to be the basis of the world's most intense UCN 
source, and with minor modifications we expect 300 UCN/cc
densities in bottles of volumes 10's of liters; the practical
limit to a spallation pulse for a target assembly
amenable to this project is 50 $\mu$C in one second; this
would be repeated every 10 sec so the average current would be
about 5 $\mu$A.  Questions relating to heating of the sample
and generation of the ortho state due to radiation remain
to be studied. Details of
the Los Alamos effort will soon be published by the collaboration.

\end{document}